\documentclass[%
 reprint,
 superscriptaddress,
 amsmath,amssymb,
 aps,
 pre,
]{revtex4-1}

\usepackage{graphicx}
\usepackage{epstopdf}
\DeclareGraphicsExtensions{.png,.pdf,.eps}
\usepackage{subfigure}
\usepackage{color}
\usepackage{tabularx}

\graphicspath{{./Fig/}}

\begin{document}

%\title{Modeling aging and deletion of social relationships in social networks}
\title{Modeling the role of relationship fading and breakup in social network formation}

\author{Yohsuke Murase}
\email{yohsuke.murase@gmail.com}
\affiliation{RIKEN Advanced Institute for Computational Science, 7-1-26, Minatojima-minami-machi, Chuo-ku, Kobe, Hyogo, 650-0047, Japan}
\affiliation{CREST, Japan Science and Technology Agency 4-1-8 Honcho, Kawaguchi, Saitama, 332-0012, Japan}
\author{Hang-Hyun Jo}
\affiliation{BK21plus Physics Division and Department of Physics, Pohang University of Science and Technology, Pohang 790-784, Republic of Korea}
\affiliation{Department of Computer Science, Aalto University School of Science, P.O. Box 15500, Espoo, Finland}
\author{J\'anos T\"or\"ok}
\affiliation{Department of Theoretical Physics, Budapest University of Technology and Economics, H-1111 Budapest, Hungary}
\affiliation{Center for Network Science, Central European University, N\'ador u. 9, H-1051 Budapest, Hungary}
\author{J\'anos Kert\'esz}
\affiliation{Center for Network Science, Central European University, N\'ador u. 9, H-1051 Budapest, Hungary}
\affiliation{Department of Theoretical Physics, Budapest University of Technology and Economics, H-1111 Budapest, Hungary}
\affiliation{Department of Computer Science, Aalto University School of Science, P.O. Box 15500, Espoo, Finland}
\author{Kimmo Kaski}
\affiliation{Department of Computer Science, Aalto University School of Science, P.O. Box 15500, Espoo, Finland}

\begin{abstract}
In social networks of human individuals, social relationships do not necessarily last forever as they can either fade gradually with time, resulting in ``link aging'', or terminate abruptly, causing ``link deletion'', as even old friendships may cease.
In this paper, we study a social network formation model where we introduce several ways by which a link termination takes place.
If we adopt the link aging, we get a more modular structure with more homogeneously distributed link weights within communities than when link deletion is used.
By investigating distributions and relations of various network characteristics, we find that the empirical findings are better reproduced with the link deletion model.
This indicates that link deletion plays a more prominent role in organizing social networks than link aging.
\end{abstract}

\date{\today}
\maketitle

\section{Introduction}\label{sec:intro}

Interdisciplinary efforts in network science have considerably deepened our understanding about the structure and dynamics of the society \cite{lazer2009computational,sen2013sociophysics}.
This active development of the field is largely attributed to the huge amount of digital information that has become available due to rapid development of the information and communication technology (ICT).
Many empirical studies have been conducted on social networks including those based on email \cite{kossinets2006empirical}, mobile phone call (MPC) \cite{onnela2007analysis,onnela2007structure,blondel2015survey}, short-message communication, social networking services (SNS) \cite{Kwak2010,Ugander2011}, and scientific collaborations \cite{newman2001structure}.
Among these, MPC data sets play a special role because mobile phones are frequently used in our daily life and the coverage of the service is almost 100\% for adults in large part of the world.
As a matter of fact, an analysis of the MPC data set \cite{onnela2007analysis,onnela2007structure} has been successful in validating the ``strength of weak ties'' hypothesis proposed by Granovetter \cite{granovetter1973strength}, stating that if the tie between two persons is strong, then the overlap between their neighbors will be large.
This has far-reaching consequences for the overall structure of the social network: It suggests that the network consists of strongly wired communities connected by weak ties.
This structure was proven at a societal level by applying percolation analysis to a social network, in which the duration of a phone call between two individuals serves as a proxy for their social tie strength.
Hereafter we will refer to these networks as having Granovetterian structure.
This led to the attempts to constructing a model of social network formation reflecting the observed structures \cite{kumpula2007emergence,jo2011emergence}.
More recently, this model was generalized to reflect the multiplex character of social networks \cite{murase2014multilayer}.

In the original model by Kumpula et al. \cite{kumpula2007emergence} two main mechanisms for the link-formation process were introduced, namely cyclic and focal closures, as they have been empirically observed to be the fundamental mechanisms for creating social ties \cite{kossinets2006empirical}.
Here the cyclic closure describes the formation of a link with the neighbor of one's network neighbor or, in other words, with a friend of a friend.
The focal closure refers to making a link with the one sharing an attribute independently of the local network topology or geodesic distance between the individuals.
In addition to these two mechanisms, a link reinforcement was introduced to correspond to the general observation that social ties get stronger when they are used.

While in the previous studies the cyclic and focal closure mechanisms have been shown to play an important role in generating a Granovetterian structure, much less is known about the role of the process in which a relationship becomes terminated.
In the original model \cite{kumpula2007emergence} nodes (and all their links) were removed with a small probability or with a slow rate and new nodes without any links were added to the system to replace them.
This way the asymptotic stationarity of the system with some average degree could be maintained.

Deletion of a node corresponds to the death of a person or giving up a service or moving far away so that all contacts break. 
In real human society, however, the termination of a relationship may occur in various ways and there is no empirical support that the deletion of a node would be the main mechanism for terminating links.
Even without a removal of a person, a relationship may end abruptly, for example, when a couple in intimate relationship suddenly break up.

According to an empirical study, an individual drops out one member of her relationships every $7.2$ months on average~\cite{dunbar2014sex}.
Apart from these rather sudden and drastic changes in human relationships, there is also a more gradual fading out of a friendship, which is typically seen when old but aging friends make less and less contact by time.
It is not clear whether such difference in tie terminations affects the emergent network properties.
In this paper, we investigate the effects of various link termination mechanisms and show that the link aging may lead to a significantly different network from the one with abrupt link termination or link deletion.
The comparison between the models with different link termination mechanisms indicates that the link deletion reproduces the empirical findings from the MPC data set better than other mechanisms, implying that a link deletion plays a  prominent role in the evolution of social networks among the link termination mechanisms.

This paper is organized as follows.
First we define the models with three different link termination mechanisms, including the original node deletion mechanism.
Then we demonstrate that modular structures are more enhanced under the link aging than other mechanisms.
Various local network features, such as degree distributions, link weight distributions, and link overlap, are investigated and compared with the MPC data set.
Finally, we present summary and discussion in the last section.

\section{Methods}\label{sec:model}
First we briefly review the original weighted social network model proposed in \cite{kumpula2007emergence}.
It considers an undirected weighted network of $N$ nodes, with links between them being updated by the following three mechanisms.
The first mechanism is \textit{local attachment} (LA), in which node $i$ chooses one of its neighbors $j$ with probability proportional to $w_{ij}$ that stands for the weight of the link between nodes $i$ and $j$.
Then node $j$ chooses one of its neighbors but $i$, say $k$, randomly with probability proportional to $w_{jk}$ and if nodes $i$ and $k$ are not connected, they are connected with probability $p_{\Delta}$ with a link of weight $w_{0}$.
In addition all the involved links increase the weights by $\delta$, whether a new link is created or not.
The second mechanism is \textit{global attachment} (GA), in which a node with no links or otherwise with probability $p_r$ makes a new link to a randomly chosen node with weight $w_{0}$.
The third mechanism is \textit{node deletion} (ND), in which a node loses all its links with probability $p_{nd}$, equivalently to replacing the node with a new isolated node.
These three processes, i.e., LA, GA, and ND, are applied to all nodes at each time step.
The initial condition is set to be a network without any links and the network reaches a statistically stationary state after a sufficient number of updates.
This model turned out to show the Granovetterian structure of the society with strongly wired communities connected by weak ties.

In this paper we extend the original model such that instead of ND, we consider two different link termination mechanisms: \textit{link deletion} (LD) and \textit{aging}.
In case of link deletion we assume an abrupt termination of links, such that each link is removed from the system with probability $p_{ld}$ at each time step.
Here we assume this probability to be independent of the link weight such that even a strong link can be suddenly terminated.
In contrast, for the link aging mechanism we assume a gradual degradation of the link weight, such that at each time step, weights of all the links are multiplied by an aging factor $f$ ($<1$), where $f$ controls the speed of aging.
If the link weight becomes less than a threshold value, $w_{th}$, the link is removed from the system.
From now on, models with ND, LD, and aging mechanisms will be called the ND model, LD model, and aging model, respectively.

\section{Results}

\subsection{Global modular structures}\label{sec:global}

\begin{figure}
\begin{center}
  \includegraphics[width=.48\textwidth]{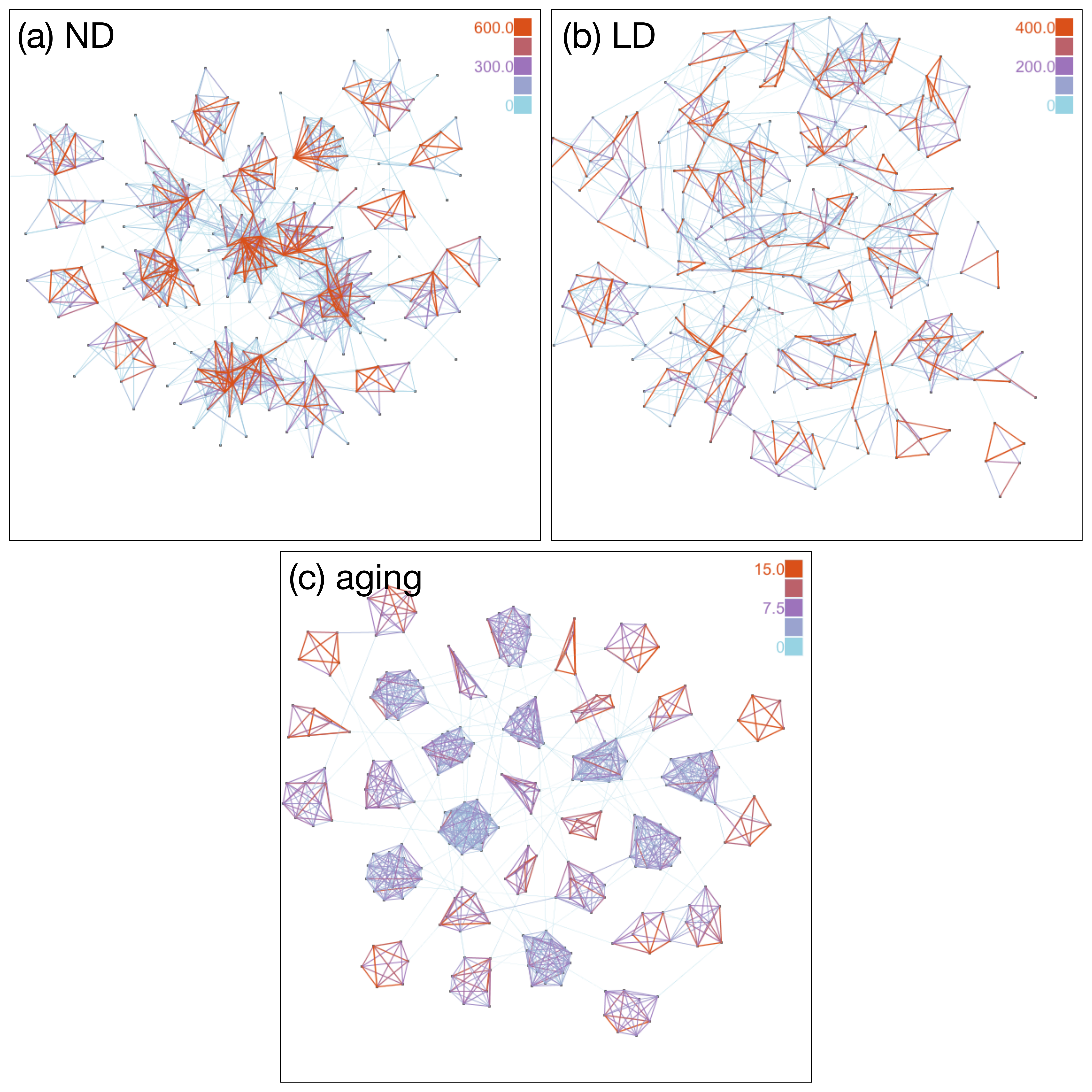}
  \caption{
    {\bf Network snapshots.}
    Snapshots of typical network structures for (a) ND, (b) LD, and (c) aging mechanisms of link removals.
    Networks are drawn by using a force-directed graph drawing algorithm.
    Parameters are controlled such that these three networks have comparable average degrees (see Table~\ref{table:degrees}).
    Color of links denotes the link weight.
    Strong and weak links are depicted in red and light blue, respectively.
    Note that different color scales are used for visibility.
    For the aging mechanism, community structures are clearly visible and the link weights in each community are more homogeneous.
    In the networks using the ND and LD mechanisms, strong and weak links coexist in each community.
  }
  \label{fig:snapshot}
\end{center}
\end{figure}

In Figure~\ref{fig:snapshot} we show for comparison typical snapshots of the networks for three different link termination mechanisms with similar average degree.
All the networks show modular structures, where the links connecting communities are typically weaker than the intra-community links, thus indicating the existence of the Granovetterian structure.
However, the strength of the modularity varies significantly.
The aging model shows clearly the strongest modularity among three networks while the LD model shows less modular structure than the others.
Another interesting finding for the aging model is that the link weights within a community are more homogeneous.
Furthermore, typical link weight in a community is inversely correlated with the community size.
This is contrasting to the other models, where each community has both weak and strong links.
In the following, we investigate these observations quantitatively.

Here we performed simulations by setting $N=10^4$, $p_{\Delta} = 0.05$, $p_r=0.0005$, $w_0 = 1$, and $\delta = 1$.
For the ND model with $p_{nd} = 0.001$ a typical Granovetterian structure is observed \cite{kumpula2007emergence} such that the average degree is $11.0$.
In order to compare the results with the LD and aging cases, we control the parameters for the LD and aging models so that average degrees are comparable with that of the ND model.
We adopted $p_{ld}=0.0035$ for LD model, and $f = 0.9$, $w_{th}=0.01$, and $p_r=0.005$ for aging model.
Note that a higher $p_r$ is used for the aging model, otherwise the fragmentation of the network occurs.
The results were obtained by running the simulations up to $25000$ time steps and averaging over $50$ realizations.
The average degrees and other quantities are summarized in Table \ref{table:degrees}.
We note that qualitatively similar results are observed for reasonable ranges of parameters other than those specified above.

\begin{table}
  \caption{
    {\bf Average degrees, clustering coefficients, and maximum modularity.}
    The average degree $\langle k \rangle$, global clustering coefficient $C$, maximum modularity for binary graph $Q$ are summarized for the ND, LD and aging models together with the simulation parameters.
    Relative fluctuations of link weights in each community, $\sigma_{c}/\langle w \rangle_{c}$, are also presented.
  }
  \label{table:degrees}
  \begin{tabular}{c|ccc}
    & ND & LD & aging \\ \hline
               & $p_r = 0.0005$ & $p_r = 0.0005$ & $p_r = 0.005$ \\
    parameters & $p_{nd} = 0.001$ & $p_{ld}=0.0035$ & $f=0.9$ \\
               & & & $w_{th}=0.01$ \\ \hline
    $\langle k \rangle$ & $11.0$ & $11.8$ & $11.1$ \\
    $C$ & $0.65$ & $0.23$ & $0.87$ \\
    %$Q_{w}$ & $0.992$ & $0.895$ & $0.994$ \\
    %$Q_{u}$ & $0.931$ & $0.564$ & $0.951$ \\
    $Q$ & $0.931$ & $0.564$ & $0.951$ \\
    $\sigma_{c}/\langle w \rangle_{c}$ & $1.87$ & $1.93$ & $0.46$
  \end{tabular}
\end{table}

One of the most prominent differences between these models is found in their community structures.
Although all models show community structure, the aging model has the most modular structure while the LD model has a lower modularity.
In Table \ref{table:degrees} we find that the aging model has the highest clustering coefficient and maximum modularity among three models.
For calculating modularity $Q$ we used the Louvain method \cite{blondel2008fast} for binary graphs, a fast and efficient method for detecting non-overlapping communities.
We find that the aging model has a slightly larger modularity than the ND model, while the LD model has a much smaller modularity than the ND model.
It should be noted that the difference between the aging and ND models is more remarkable than one might think because ten times larger $p_r$ value is used for the aging model.
If in contrast we use the same $p_r$ for the aging model, we get $Q > 0.99$ and the network is fragmented into communities most of which are cliques.
This indicates that the link aging makes the network more modular than ND or LD.
We have obtained the modularity using the Louvain method for weighted graphs to find less difference in modularities due to the fact that inter-community links are typically weaker than intra-community links.

The higher modularity for the aging model is closely related to the Granovetterian structure.
Since only weak links may be removed under the aging mechanism, inter-community links, which are typically weak, are removed more frequently.
On the other hand, links within a community do not disappear easily as their weights are strong and once the Granovetterian community structure is generated, the modular structure is enhanced even more by aging such that the communities are more persistent.
In contrast, since the ND and LD models remove links irrespective of the link weights, the intra-community strong links may be removed by chance, which results in less modular structure for the ND and LD models.

We also note the difference between the ND and LD models.
Although these two models assume an abrupt termination of the links, the ND model has higher values of clustering coefficient and modularity.
This is because LD tends to reduce the fraction of intra-community links, which becomes evident by considering a clique of size $n$ that after an ND event has $n-1$ links removed and a clique of size $n-1$ remains, meaning that the network has still a high modularity.
However, if we remove $n-1$ links randomly by link deletion, the remaining network is no longer a clique, and we get a lower modularity because the number of links in a community is decreased while the number of nodes remains same.
This implies that the community structure is in general more robustly maintained for the ND model than the LD model.

Another notable difference of the aging model is the homogeneity of the link weights within each community.
The link weights in a community are approximately similar for the aging model whereas the networks for the ND and LD models contain various weights of links in each community.
In order to quantify this homogeneity, we introduced the relative fluctuation of intra-community link weights, which is defined as follows.
First we identified communities then, for each community, calculated the relative fluctuation $\sigma_{c} / \langle w \rangle_{c}$, where $\sigma_{c}$ and $\langle w \rangle_{c}$ are the standard deviation and the average of the intra-community link weights, respectively.
In order to distinguish the inter-community and the intra-community links, we needed to identify the non-overlapping communities.
As shown in the Table~\ref{table:degrees}, the aging model has a $\sigma_{c} / \langle w \rangle_{c}$ value less than one, indicating a smaller variation in link weights than observed in the other models.

\begin{figure}
\begin{center}
  \includegraphics[width=.4\textwidth]{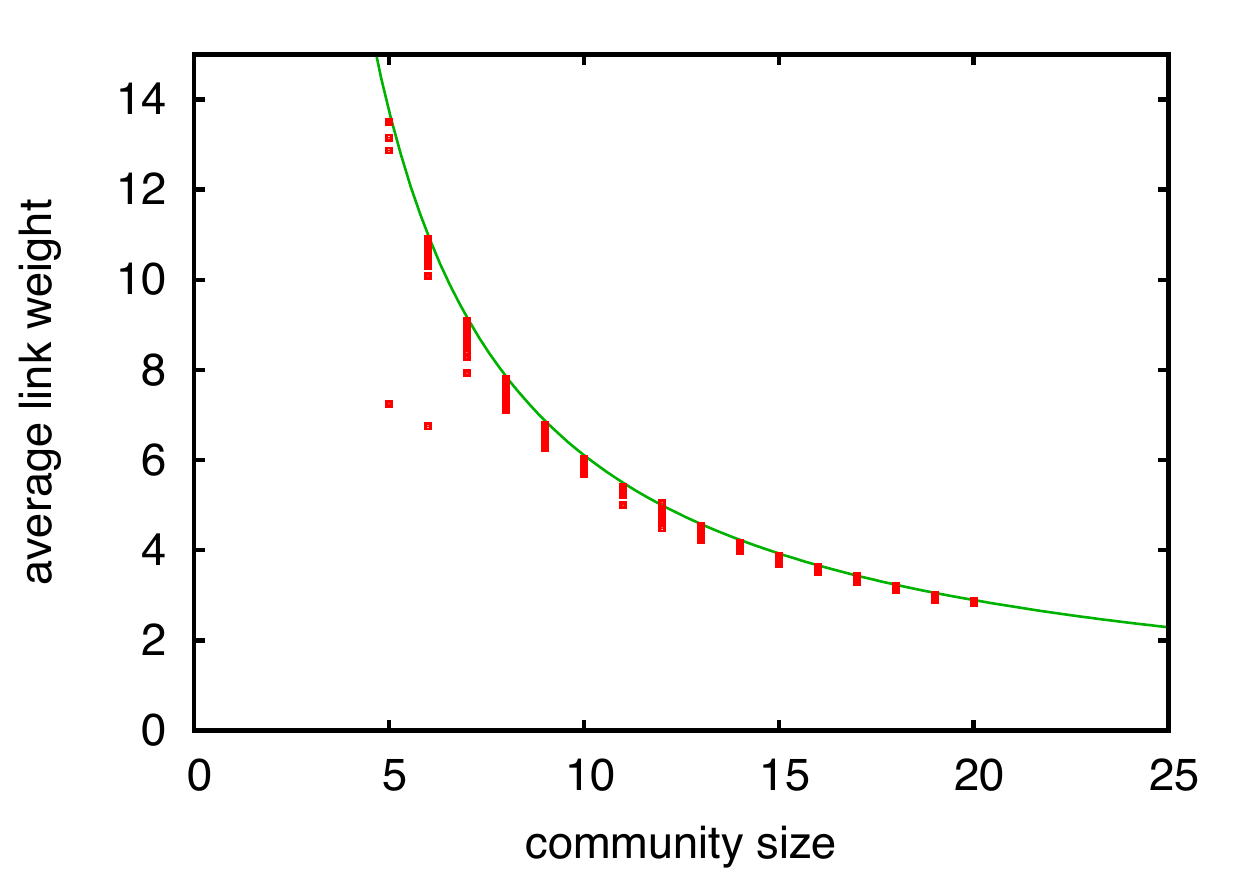}
  \caption{
    {\bf Average link weight of intra-community link as a function of the community size.}
    Communities are obtained from a single run for the aging model and each point corresponds to a community.
    A fitting function proportional to $1/(n-1)$ is also shown.
  }
  \label{fig:community_link_weight}
\end{center}
\end{figure}

It is also notable that the average link weight of a community $\langle w \rangle_{c}$ is inversely proportional to the community size.
Figure~\ref{fig:community_link_weight} shows the average link weight as a function of the community sizes for a single run of the aging model\footnote{
For this plot, we used an intermediate solution obtained by the Louvain method because the final solution does not show a clear convergence of the data points to a single curve but to a few curves.
It is probably because multiple communities are merged in the final solution due to the resolution limit \cite{fortunato2007resolution}.
}.
It clearly shows that the average link weight decreases as the community size increases, and it is well fitted by a function proportional to $1/(n-1)$.
This relation between the average link weights and the community size for the aging model can be explained using an approximation that a community is a clique of size $n$ having a uniform link weight $\langle w \rangle$.
This is a reasonable approximation because the modularity of the simulated network is close to one.
The link weight $\langle w \rangle$ is estimated by the balance between the reinforcement by LA and the loss by aging.
Here, we note that the increase of the link weights by GA is negligibly small compared to the reinforcements by LA.
The probability $p$ that a link $l$ is included in one LA event is given by the number of possible triangles including $l$ divided by the number of possible triangles in the clique,
\begin{equation}
p = \frac{n-2}{\binom{n}{3}} = \frac{6}{n(n-1)}.
\end{equation}
At each time step, the expected increase of a link weight is $pn$ because LA happens $n$ times in that clique each time step.
On the other hand, the decrease of a link weight is $\langle w \rangle (1-f)$.
Comparison of the increase and the decrease gives
\begin{equation}
\label{eq:w_eq}
  \langle w \rangle = \frac{6}{(1-f)(n-1)},
\end{equation}
which is in a quite good agreement with Fig.~\ref{fig:community_link_weight}.

Aging is a multiplicative process, which dominates strong links, while reinforcement is additive with major impact on weak links.
The link weights in a community tend to converge to their equilibrium value $\langle w \rangle$. 
Thus, the dynamics of $w$ is stable around its equilibrium value.
A more detailed linear stability analysis is given in Appendix~\ref{sec:linear_stability}.

\begin{figure}
\begin{center}
  \includegraphics[width=.5\textwidth]{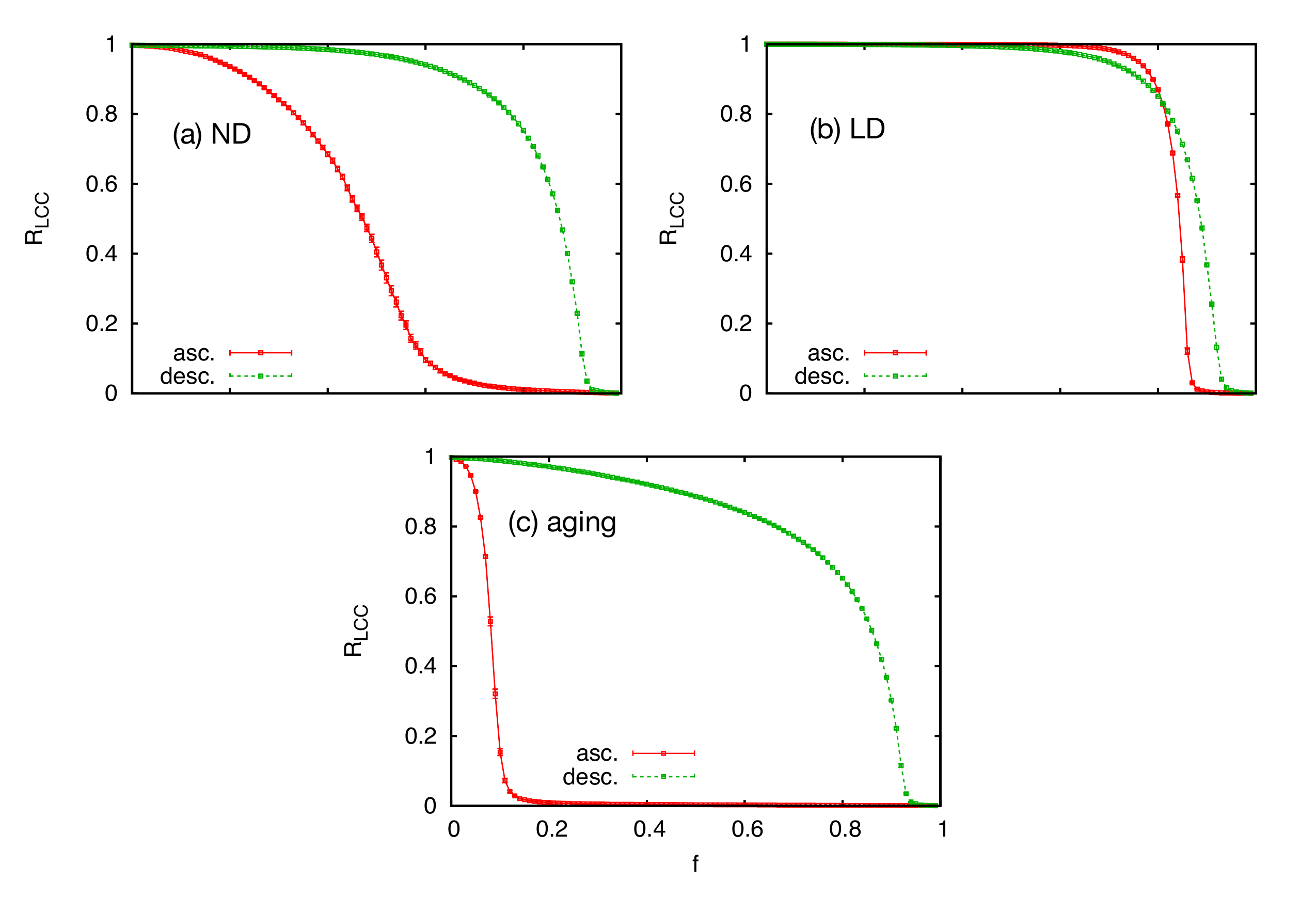}
  \caption{
    {\bf Link percolation analysis.}
    Link percolation analysis for the (a) ND, (b) LD, and (c) aging models.
    We plot the relative size of the largest connected component, $R_{\rm LCC}$, as a function of the fraction of removed links $f$.
    Red solid (green dashed) curve corresponds to the case when links are removed in ascending (descending) order of link weights.
    The error bars show standard errors.
  }
\label{fig:percolation}
\end{center}
\end{figure}

The differences between the community structures are also observed in the percolation analysis.
Provided that links are sorted according to their weight, removing the links in the ascending order results in a sharp transition at a relatively early stage indicating the fragmentation of the society.
In the opposite case, when links were eliminated in the descending order, the percolation threshold is significantly higher than that for the ascending order because strong links are within the communities.
Figure~\ref{fig:percolation} shows that all models have the Granovetterian structure that a difference between ascending and descending order percolation thresholds, $\Delta f_c$, is observed.
Although they show the Granovetterian structure similarly, the amount of $\Delta f_{c}$ shows a large variation.
The aging model shows a much lower percolation threshold for the ascending order compared to the ND and LD models, while the percolation thresholds for the descending order are comparable.
This is another indicator of the high modularity for the aging model.
Since the fraction of the inter-community links to the intra-community links is low when the network is modular, removal of smaller number of weak links may lead to the fragmentation of the network.
Furthermore, as we have seen, weak links are present in the communities for the ND and LD models, which makes the percolation threshold for the ascending order even higher.
Thus, the aging model shows a faster fragmentation when weak links are removed first, while the LD model remains connected up to a higher percolation threshold.

\subsection*{Local network properties}\label{sec:local}

Not only the global modular structures but also several local network properties show qualitative differences between the three models.
In this Section, we obtain several local network properties and compare them with those of the MPC data set.

\begin{figure}
\begin{center}
  \includegraphics[width=.5\textwidth]{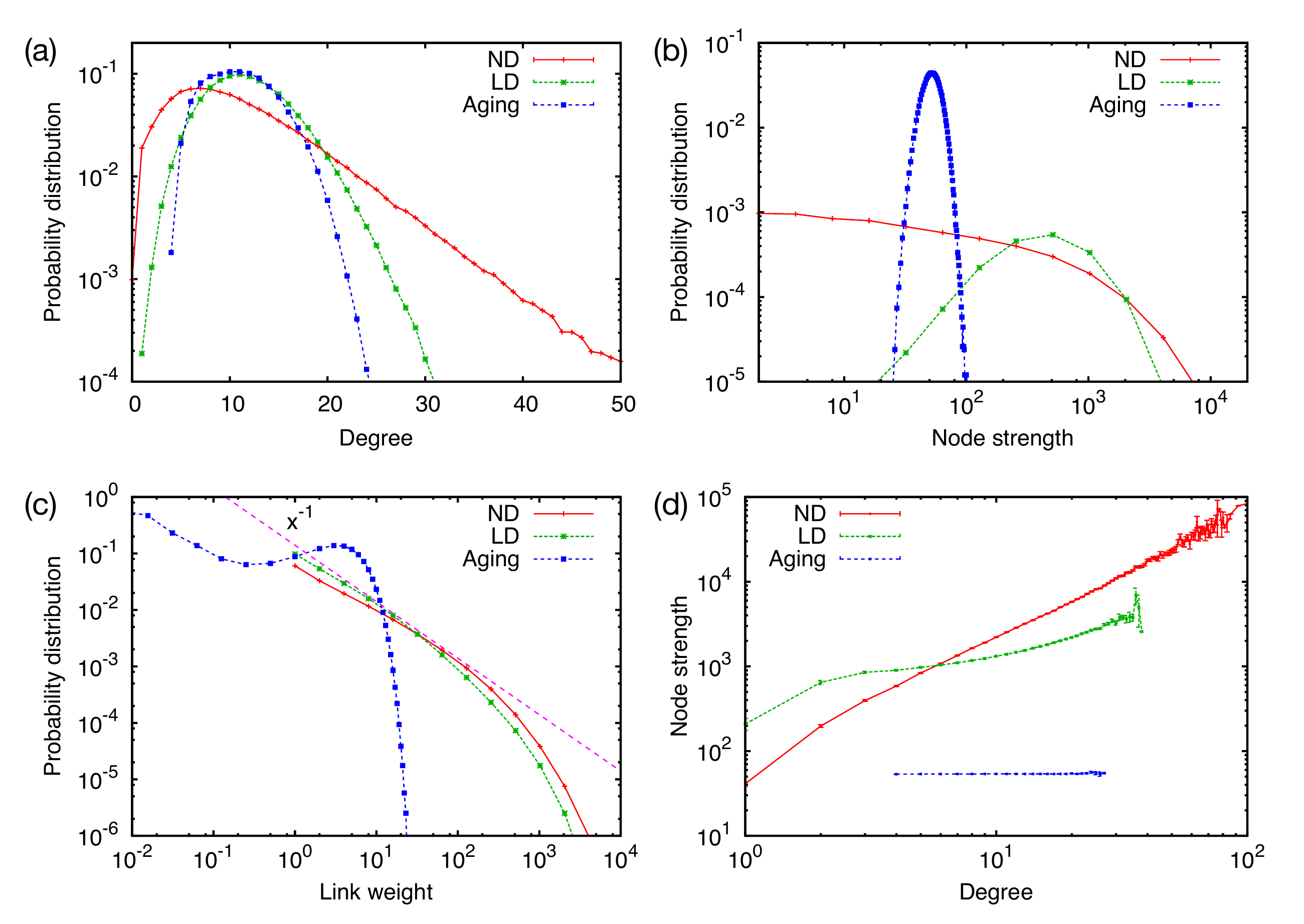}
  \caption{
    {\bf Local network properties.}
    (a) Degree distributions, (b) node strength distributions, (c) link weight distributions, and (d) correlations between node strength and degree for the ND, LD, and aging models.
    In (c), a fitting line of $x^{-1}$ is also shown.
  }
  \label{fig:degree_distribution}
\end{center}
\end{figure}

Degree distributions for three models are shown in Fig.~\ref{fig:degree_distribution}(a). While the LD and aging models show Gaussian-like distributions, the ND model has an exponential tail for large $k$ values. This is because a node in the ND model has monotonically increasing degree for most of the time until it is deleted. On the other hand, in the other models degrees of nodes can increase and decrease. Thus, the degree fluctuates around its mean value and the distribution gets similar to a Gaussian distribution due to the central limit theorem. In Fig.~\ref{fig:degree_distribution}(b) we depict the distributions of the node strength, defined as the sum of the link weights a node has. As in the case of degree distributions we find that the LD and aging models show peaks while the ND model shows a monotonically decaying behavior. In comparison, the MPC and many other empirical data sets show a monotonically decreasing degree and strength distributions \cite{onnela2007analysis}. Hence, the ND model seems to reflect best the empirical findings on that dataset. However, the monotonically decaying behavior might not correspond to common sense that in reality a person usually has a typical number of friends and typical amount of communication \cite{dunbar1992neocortex}. Thus we would expect that the monotonically decaying behavior found in the data set is due to its incompleteness and would change if other channels of communication could be taken into account. For this reason we speculate that the LD and aging models might capture the reality better than the ND model.

Figure~\ref{fig:degree_distribution}(c) shows the link weight distributions, where we find clear differences between the models. Both the ND and LD models show a monotonically decaying behavior which is approximated by a power law with an exponent close to $-1$. It indicates the coexistence of strong and weak links. In contrast, the aging model shows an initial power law decay for $w < 1$, but has a peak at around $w \approx 5$ and then decays quickly. 
As the MPC data set shows monotonically decaying behavior, we conclude that the ND and LD models fit better with the empirical data than the aging model.

The initial power law decay in the link weight distribution for the ND and LD models is explained as follows. A link weight increases by the reinforcements of LA.
Assuming that the probability that a link is selected by LA is proportional to its link weight, then the dynamics of the link weight $w$ is approximated as $dw/dt = c{w}$, where $c$ is a constant. Hence the expected weight of a link at age $t$ is $w(t) = \exp{(t/\tau_{r})}$ since $w(0)=1$, where $\tau_{r}$ is a constant characterizing the speed of the reinforcements of the link weight. Since a link is removed randomly with a constant rate, the age distribution of links is an exponential distribution $P(t) = \exp{(-t/\tau_{d})}/\tau_{d}$, where $\tau_{d}$ is a time constant of the removal process. Using these equations, $P(w) \propto w^{-(1 + \tau_{r}/\tau_{d})}$ is obtained and its exponent is close to $-1$ since $\tau_{r} \ll \tau_{d}$, i.e., the reinforcements occur much more frequently than the removal of a link. This is consistent with the simulation results when $w$ is small. The deviation from power-law behavior is observed for large $w$ because of natural cutoffs due to node and link deletions.

On the other hand the initial power law decay in the link weight distribution for the aging model is explained by the decaying dynamics of $w$. Most of the links whose weight is less than one consist of the links which have not been reinforced after they are generated. The time evolution of the weight of such a link is given by $w(t) = \exp{(-t/\tau_{a})}$, where $\tau_{a}=1/(1-f)$ is a time constant for aging. Assuming a new link is generated by a constant rate, the distribution of the link age becomes uniform.
With this assumption we get the link weight distribution proportional to $w^{-1}$. Although the initially decaying behavior of the distribution is similar to those for the ND and LD models, the underlying mechanism is different.

The node strength for the aging model does not show a dependence on the degree in Fig.~\ref{fig:degree_distribution}(d), as the strength is given as $s=\langle w\rangle (n-1)=6/(1-f)$. This is clearly different from results of the ND and LD models and the MPC data set, where a node of a higher degree has a higher node strength.

\begin{figure}
\begin{center}
  \includegraphics[width=.5\textwidth]{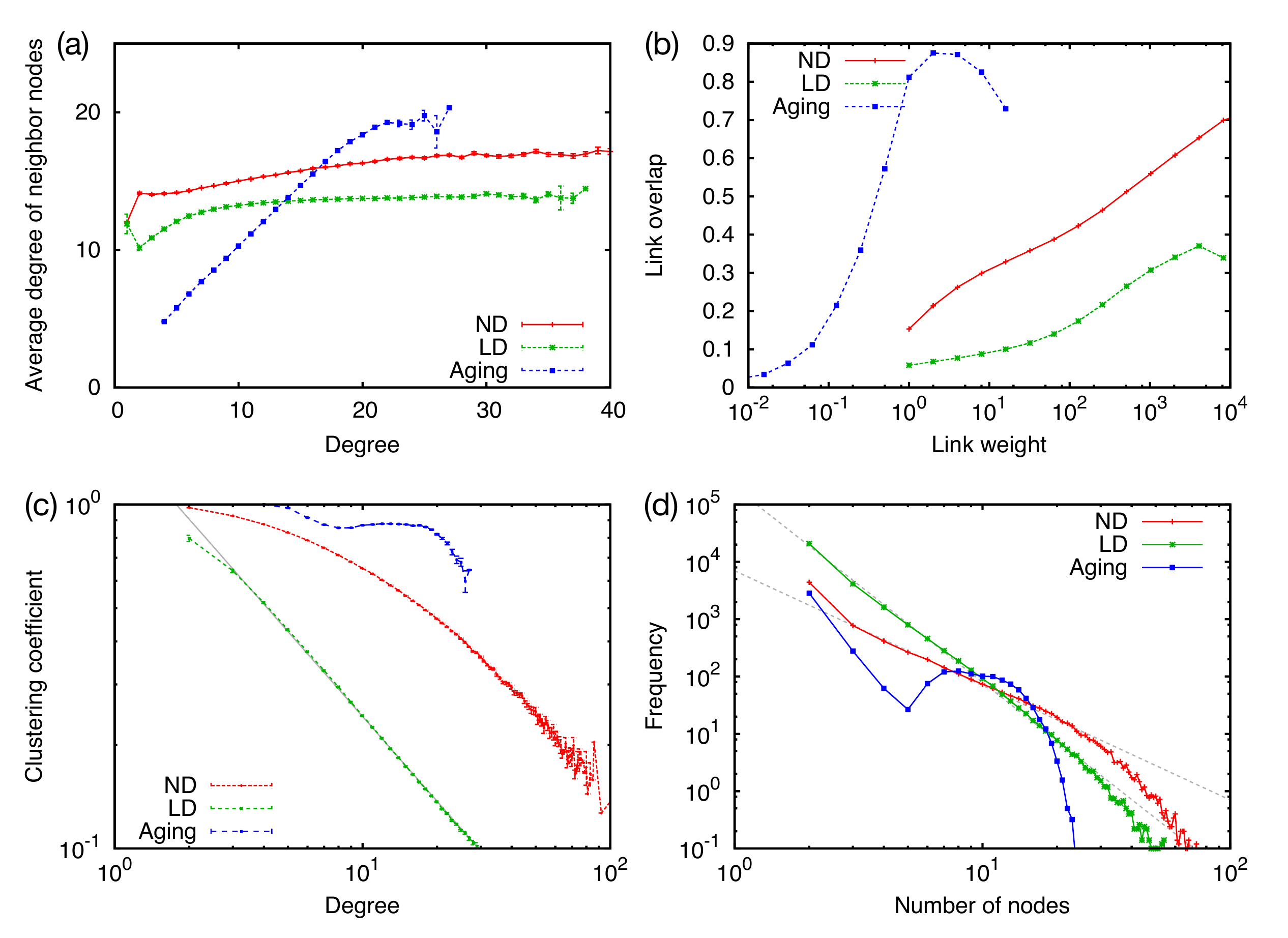}
  \caption{
    {\bf Relations of various network characteristics.}
    (a) Average degree of neighboring nodes as a function of degree, (b) link overlap as a function of link weight, (c) local clustering coefficient as a function of degree, and (d) community size distribution for the ND, LD, and aging models. In (c) a fitting line corresponding to $x^{-0.82}$ is also shown. In (d) communities are identified using the link community detection method. Fitting by power laws with exponent $-3.4$ and $-2$ are shown as guides to the eyes.
  }
  \label{fig:neighbor_degree_distribution}
\end{center}
\end{figure}

The observations from MPC and other social network data sets show assortative mixing, i.e., people having many friends tend to be connected to those who also have many friends. This is measured by the correlation of degrees between neighboring nodes. Figure~\ref{fig:neighbor_degree_distribution}(a) shows the average degree of the nearest neighbors of a node with degree $k$. We find assortative mixing for all three models although the degree of the correlation is different. The strong positive correlation for the aging model arises from the strong modular structure. Since a node and its neighbor often belong to the same community and the communities are often cliques, nodes in the same community have similar degrees, which is determined by the size of the community. The networks for the ND and LD models are less modular, thus the positive correlation is not strong as the aging model.

One can ascertain the Granovetterian structure also by correlating the link weight and its overlap measure $O_{ij}$. The topological overlap of the neighborhood of two connected nodes $i$ and $j$ is defined by the fraction of common neighbors \cite{onnela2007structure}:
\begin{equation}
  O_{ij} = \frac{n_{ij}}{(k_i-1)+(k_j-1)-n_{ij}},
\end{equation}
where $n_{ij}$ is the number of neighbors common to both nodes $i$ and $j$. Networks having the Granovetterian structure show an increasing behavior of $O_{ij}$ according to $w$. As shown in Fig.~\ref{fig:neighbor_degree_distribution}(b), positive correlations are found for the ND, LD, and aging models, while for the aging model it is found for $w<1$. This indicates the existence of the Granovetterian structure. This positive correlation supports the robustness of the Granovetterian structure with respect to the modifications of link termination mechanism.

The negative correlation found in the overlap for the aging model for $w>1$ is explained by the negative relation between the community size and the link weight. As we have seen in the previous section, most links with weight larger than one are within communities and the link weight is inversely related to the community size. Links within larger communities usually have larger overlap. This is because the number of common neighbors is approximately proportional to the community size whereas the number of inter-community links do not depend on the community size since most of them are created by GA mechanism, which is independent of the network topology. Therefore, weaker intra-community links, which often belong to a larger community, tend to have a larger overlap. Even though $O_{ij}$ and $w$ are negatively correlated for intra-community links, the Granovetterian picture that communities with strong ties are connected by weak ties is still valid.

Empirical MPC data set has also shown a non-monotonic $O(w)$ as observed in the aging model. While most of the links show a positive correlation between $w$ and $O(w)$, the top five percent links of link weights show a negative correlation \cite{onnela2007analysis}. However, the underlying mechanism of MPC data set is not explained by the aging model. This is because the negative correlation for the aging model is originated from the negative correlation between the average degree and the link weights. In the MPC data set, however, the correlation between weight and degree product $\langle w_{ij} | k_i k_j \rangle$ is negligible \cite{onnela2007analysis}, which is not present in the aging model. In the case of the MPC data, the change in the trend in the $O_{ij}$ vs $w_{ij}$ plot could be traced back to a different origin: The extremely high values of weights were results of calls between persons, who almost exclusively talked to each other \cite{onnela2007structure}. 

Figure~\ref{fig:neighbor_degree_distribution}(c) displays the relation between the local clustering coefficient and the degree. While the LD and ND models show a smoothly decreasing behavior, the aging model has a plateau around at $k=10$. Not only MPC but also other social network data sets \cite{saramaki2007generalizations} often show negative correlations and the curve is often fitted by a power law decay with exponent close to $-1$. Therefore, the aging model does not on its own explain the empirical data. The reason could be that the aging model produces a rather narrow degree distribution where the asymptotics cannot develop.

The community size distribution is yet another interesting statistics. Ahn \textit{et al.} have demonstrated that the MPC data set shows a power law with exponent $-3$ in the size distribution of link communities \cite{ahn2010link}. In order to compare this empirical analysis with our models, we calculate the community size distribution using the link community detection method. In Fig.~\ref{fig:neighbor_degree_distribution}(d) we show the distributions of the number of nodes in a link community, where we have adopted the method for a weighted graph. We observe that the ND and LD models show approximate power-law decaying behavior, which is qualitatively consistent with the observation from the MPC data set. However, the aging model has a clearly different profile as it shows an initial power law decay and a peak indicating the existence of a characteristic size. If we call the modules shown in Fig.~\ref{fig:snapshot}(c) as ``node community'', the former region generally corresponds to the links connecting node communities and the latter corresponds to the links within node communities. This distinction between inside and outside node communities is detectable only for the aging model. We speculate that this is because the modular structure for the aging model is so strongly emphasized that the distinction becomes visible. The ND and LD models as well as the MPC data set do not show distinct regions, implying that the aging model over-emphasizes the community structure.

The statistics investigated above together with the MPC data set observations are summarized in Table~\ref{table:stat_summary}.
In the table, the expected behavior which matches to our common sense is also shown since we suspect some of the MPC data set does not reflect the real society.
This is not only because of the limited observation period but because of other communication channels often used as well as MPC in our daily life.
As discussed in \cite{dunbar1992neocortex}, we expect there is a typical degree or amount of communications therefore degree distribution or node strength distribution should have a peak.
We also expect the link overlap generally have a positive relationship with link weights \cite{granovetter1973strength}.
These expectations are shown in the last column of the table although these are debatable.
We find that the ND and LD models reproduce reasonably well most of the general trends of the MPC data set and the common sense while the aging model, at least on its own, does not do that.
Hence we conclude that overall the MPC data are quite successfully reproduced by the ND or LD models.

\begin{table*}
  \caption{
    {\bf Summary of local network properties for the ND, LD, and aging models.}
    The arrows indicate the general trend of the profile: $\nearrow$ ($\searrow$) implies that the profile is monotonically increasing (decreasing). If the curve is initially increasing and then decreasing, it is denoted as $\nearrow \searrow$.
    \label{table:stat_summary}
  }
  \begin{tabular}{c|ccc|c|c}
    & ND & LD & aging & MPC & common sense \\ \hline
    degree distribution & $\nearrow \searrow$ & $\nearrow \searrow$ & $\nearrow \searrow$ & $\searrow$ & $\nearrow \searrow$ \\
    node strength distribution & $\searrow$ & $\nearrow \searrow$ & $\nearrow \searrow$ & $\searrow$ & $\nearrow \searrow$ \\
    link weight distribution & $\searrow$ & $\searrow$ & $\nearrow \searrow$ & $\searrow$ & $\searrow$ \\
    strength as a function of degree & $\nearrow$ & $\nearrow$ & $\rightarrow$ & $\nearrow$ & $\nearrow$ \\
    average degree of neighbors as a function of degree & $\nearrow$ & $\nearrow$ & $\nearrow$ & $\nearrow$ & $\nearrow$ \\
    link overlap as a function of link weight & $\nearrow$ & $\nearrow$ & $\nearrow \searrow$ & $\nearrow \searrow$ & $\nearrow$ \\
    local clustering coefficient as a function of degree & $\searrow$ & $\searrow$ & $\rightarrow$ & $\searrow$ & $\searrow$ \\
    link community size distribution & $\searrow$ & $\searrow$ & $\searrow \nearrow \searrow$ & $\searrow$ & $\searrow$
  \end{tabular}
\end{table*}

\section{Discussion}\label{sec:summary}

In this paper, we have investigated how the mechanism of link termination affects the emerging network structures by introducing and comparing three different models, namely node deletion (ND), link deletion (LD), and aging models. First of all, we would like to emphasize that the Granovetterian structure is robustly observed for all three models. It supports the observation by the previous studies that cyclic and focal closures play key roles in generating the Granovetterian structure.

On the other hand, there are significant differences between the ND, LD and aging models. We find that aging promotes modular structure in networks having the Granovetterian structure. This is because inter-community weak links are often deleted while strong links existing within communities need a longer time to be removed, thus making the communities more persistent under the aging than under ND or LD. It is also notable that the aging makes link weights in each community more homogeneous. In contrast, with link deletions assumed in the ND or LD models communities are less modular and links have variable weights.
With the LD model it is even harder to maintain high modularity than with the ND model because the density of links in a community tends to decrease more rapidly with LD.

Local network properties for these three models, such as degree distribution, node strength distribution, and link overlaps, have also been investigated and compared with the MPC data set observations. The results show that the ND and LD models reproduce the MPC data set while the aging model shows different profiles due to its highly modular structure with homogeneous links. This implies that abrupt link deletion is the major mechanism for deleting ties in real society, while link aging, being presumably present, plays a minor role.    

The difference between the ND and LD models are less clear, but we think that the LD model describes the reality better than the ND model. Degree and node strength distributions for the LD model do not show a monotonically decaying behavior but have peaks at characteristic scales. This matches better to our common sense that each person has some amount of resources devoted to communication although we do not have conclusive evidence from empirical data yet. Furthermore, an empirical analysis has shown that the relationships terminate with a significant rate even with individuals who have had a long term relationship \cite{dunbar2014sex}. Thus, the LD model would be a good starting point for a simple model of social networks.

In this paper, we have focused on non-overlapping communities but in reality, the communities may be overlapping as discussed in \cite{ahn2010link}. In order to realize such overlapping communities, we have recently proposed a multilayer model, where the network in each layer is generated in a similar way to the original weighted social network model \cite{murase2014multilayer}. In that case the node deletion takes place in each layer independently, which corresponds to an event that a node leaves one of its communities. This may also sound a reasonable assumption since we sometimes lose links not only to a person but collectively to his or her close friends. Although we expect most of the properties investigated in this paper to hold for the multilayer model, further studies on multilayer models would be interesting.

In reality, the link termination process happens in more complicated and various ways and our model is at best a simplification. Although we assume that the link does not recover once it is removed, more than half of the real conflicts are reconciled \cite{dunbar2014sex}. Furthermore, the probability of link deletion should also depend on the age of the friendship \cite{burt2000decay}. In Twitter data set, it was observed that ``unfollowing'' behavior depends on the local network topology such as reciprocity or existence of common friends \cite{xu2013structures}. As these aspects are missing in our models the future investigations on these factors are expected to make the model even more realistic.

Two aspects should be emphasized finally. First, when we compare our models with data, we have to restrict ourselves to the available sources, which means single channel communication records. This is an obvious source of bias, as multiple channels are crucial in human communication, which may change some of the statistics qualitatively. Second, the three different mechanisms separately treated here in disjunct models are simultaneously present in a real human society. The advantage of the present study is that we have gotten insight about their role in the formation of the social network.

\begin{acknowledgments}
Y.~M. appreciates hospitality at Aalto University.
H.-H.~J. acknowledges financial support by Aalto University postdoctoral program and by Mid-career Researcher Program through the National Research Foundation of Korea (NRF) grant funded by the Ministry of Science, ICT and Future Planning (2014030018), and Basic Science Research Program through the National Research Foundation of Korea (NRF) grant funded by the Ministry of Science, ICT and Future Planning (2014046922).
J.~T. and J.~K. acknowledge support from EU Grant No. FP7 317532 (MULTIPLEX).
The systematic simulations in this study were assisted by OACIS \cite{murase2014tool}.
\end{acknowledgments}

\appendix
\section{Linear stability analysis of the aging model}\label{sec:linear_stability}

In this Appendix, a linear stability analysis of the link weight $w$ around the equilibrium value $\langle w \rangle$ for the aging model is demonstrated. In order to evaluate the linear stability of $w$ around $\langle w \rangle$, we consider the case that one of the links, say the link between nodes $i$ and $j$, is stronger than other links by $\Delta w$ while other links have the same weight as $\langle w \rangle$. The expected amount of reinforcement on the link $ij$ in one time step, $\langle w_{+} \rangle$, is given by
\begin{multline}
  \langle w_{+} \rangle =
  2 \cdot \frac{w + \Delta w}{w(n-1) + \Delta w} \\
  { } + (n-2) \cdot \frac{2}{n-1} \cdot \frac{w+\Delta w}{(n-2)w+\Delta w} \\
  { } + 2 \cdot \frac{(n-2)w}{w(n-1) + \Delta w} \cdot \frac{1}{n-2}.
\end{multline}
The first, second, and third terms correspond to the case that the link $ij$ is selected as the first, second, and third link in one LA event, respectively. LA must start from the node $i$ or $j$ when the link $ij$ is selected either as the first or the third link, which gives $2$ in the first and the third terms. When the link $ij$ is selected as the second link, the LA event must start from the node except for $i$ and $j$, which yields the factor $n-2$ in the second term. If we assume $\Delta w \ll 1$ and ignore the higher order terms of $\Delta w$, the expected increase $\langle w_{+} \rangle$ is calculated as
\begin{multline}
  \langle w_{+} \rangle \approx
  \frac{6}{n-1} + \frac{2\Delta w}{w(n-1)} \left( 2 - \frac{2}{n-1} + \frac{1}{n-2} \right).
\end{multline}
On the other hand, the expected amount of the decrease due to aging, $\langle w_{-} \rangle$, is $(w+\Delta w)(1-f)$. When $w = \langle w \rangle$, the total change of the link weight is calculated using Eq.~(\ref{eq:w_eq}) as
\begin{equation}
  \langle w_{+} \rangle - \langle w_{-} \rangle = \Delta w (1-f) \cdot \frac{-n^2+3}{3(n-1)(n-2)},
\end{equation}
which always has a negative sign of $\Delta w$ for $n \geq 3$, indicating the dynamics of $\Delta w$ is linearly stable around its equilibrium value $\langle w \rangle$.
This demonstrates that the link weight within communities gets closer to its equilibrium point with time. Since the equilibrium link weight is shared in a community, the link weights in a community become homogeneous.

% \bibliography{complex-network}

\end{document}